\documentclass[12pt,preprint]{aastex}

\begin{document}
\title{Toward Mapping the Detailed Density Structure of Classical Be Circumstellar Disks}   

\author{J.P. Wisniewski\altaffilmark{1,2,3}, A.F. Kowalski\altaffilmark{4,5}, K.S. Bjorkman\altaffilmark{3,6}, J.E. Bjorkman\altaffilmark{3,6}, \& A.C. Carciofi\altaffilmark{7}}   

\altaffiltext{1}{NASA Goddard Space Flight Center Code 667 Greenbelt, MD 20771,  jwisnie@milkyway.gsfc.nasa.gov}  
\altaffiltext{2}{NPP Fellow}
\altaffiltext{3}{Visiting Astronomer at the Infrared Telescope Facility, which is operated by the 
University of Hawaii under Cooperative Agreement no. NCC 5-538 with the National Aeronautics and 
Space Administration, Science Mission Directorate, Planetary Astronomy Program}
\altaffiltext{4}{Department of Physics, University of Chicago 5640 S. Ellis Ave. Chicago, IL 60637}
\altaffiltext{5}{Current address: Department of Astronomy, University of Washington
Box 351580 Seattle, WA 98195, kowalski@astro.washington.edu}
\altaffiltext{6}{Department of Physics \& Astronomy, University of Toledo MS113, 
Toledo, OH 43606 USA, karen@astro.utoledo.edu, jon@physics.utoledo.edu}
\altaffiltext{7}{Instituto de Astronomia, Geof\`isica e Ci\^encias Atmosf\`ericas, Universidade de S\~ao Paulo, Rua do Mat\~ao 1226, Cidade Universit\`aria, S\~ao Paulo, SP 05508-900, Brazil, carciofi@usp.br}

\begin{abstract} The first results from a near-contemporaneous optical and infrared 
spectroscopic observing program designed to probe the detailed density structure of 
classical Be circumstellar disks are presented.  We report the discovery of 
asymmetrical infrared emission lines of He I, O I, Fe II, and the Brackett, Paschen, and 
Pfund series lines of H I 
which exhibit an opposite V/R orientation (V $>$ R) to that observed for the 
optical Balmer H$\alpha$ line (V $<$ R) in the classical Be star $\zeta$ Tau.  We interpret 
these data as evidence that the density wave which characterizes $\zeta$ Tau's disk has 
a significantly different average azimuthal morphology in the inner disk region as 
compared to the outer disk region.  A follow-up multi-wavelength observational campaign to trace the 
temporal evolution of these line profile morphologies, along with detailed theoretical modeling, is
suggested to test this hypothesis.  \end{abstract}

\keywords{stars: emission-line, Be --- circumstellar matter --- stars: individual ($\zeta$ Tau) ---
stars: individual (48 Librae)}

\section{Introduction}

Both observational and theoretical evidence indicates that classical Be stars 
are near main sequence B-type stars with gaseous circumstellar disks (e.g. see review by
\citealt{por03}); however, the mechanism(s) responsible 
for creating these disks are not well understood.  
Optical spectroscopic studies have uncovered a myriad of different line profile morphologies 
commonly observed in classical Be stars \citep{han96}, including  
double-peaked asymmetrical emission lines \citep{mcl61,dac92,han95}.  The ratio of the violet (V) to red (R) flux 
within individual lines, hereafter referred to as a V/R ratio, is observed to vary on quasi-periodic timescales of several 
to ten years, as documented for a large set of Be stars by \citet{oka97}.  While numerous theories have 
been proposed to explain this observed phenomenon, as summarized within 
\citet{han95}, these features are most  commonly interpreted as evidence of one-armed density waves within these circumstellar 
disks \citep{oka91,han95,oka97}.  Note however that the presence of phase lags between 
polarimetric and H$\alpha$ observations
has led \citet{mcd00} to suggest these waves might be more spiral-like than previously thought.  Interferometric studies \citep{vak98,ber99} have also provided evidence that Be disks deviate from 
axisymmetry, namely they have provided evidence of prograde one-armed oscillations in several disks. 

It is not uncommon to observe multiple emission lines in a spectrum of a classical Be star that 
exhibit a V/R asymmetry, and with few exceptions the same orientation of this asymmetry is observed in 
all lines.  \citet{baa85} summarizes several of these exceptions, in which a phase lag appears
to be present between optical hydrogen and Fe II lines \citep{sle82} and between different lines in the 
hydrogen Balmer series \citep{kog84}.  \citet{baa85} also suggested that a phase 
lag exists between the H$\alpha$ and hydrogen Pa 15-18 lines of $\gamma$ Cas presented by \citet{cha83}; 
however, we believe this suggestion is rather dubious given the poor resolution of these data 
and clear evidence of contamination from nearby lines.  

Few moderate resolution infrared (IR) spectroscopic studies of classical Be stars have been performed to 
date.  \citet{cla00} and \citet{ste01} have investigated a large set of Be stars in the H and K band at 
sufficient resolution to begin to search for detailed line 
profile effects.  The first tentative evidence of IR V/R reversals, analogous to the optical V/R phase lags 
summarized by \citet{baa85}, was presented in \citet{cla00}, who noted that 
the Brackett $\gamma$ and He I 2.058$\mu$m lines had opposite orientations in one of their targets.  

In this Letter, we present the initial results from a near-contemporaneous, optical and IR spectroscopic 
observing campaign designed to provide insight into the detailed density structure of classical Be circumstellar 
disks, hence diagnose the mechanisms responsible for creating these disks.  Detailed 
modeling of these data, as well as 
follow-up observations, will be presented in subsequent papers.  In Section 2 we outline the observational data to be 
discussed in this work.  An analysis and discussion of these data are presented in Section 3.  Finally, a summary of our 
main results and suggested follow-up work is presented in Section 4.

\section{Observations}

The infrared (IR) spectroscopic data presented in this paper were obtained at the 3m 
NASA Infrared Telescope Facility (IRTF), 
using the SpeX medium resolution spectrograph \citep{ray03}.  We used a 0.3 x 15 arc-second slit to 
provide wavelength coverage from 0.8-2.4$\mu$m at R $\sim$2000.  Observations of A0V stars were obtained at $\sim$15 minute intervals throughout our run, at a similar hour angle and airmass to our science targets, to 
serve as telluric standards.  Further details regarding the telluric correction 
method commonly applied to SpeX data are presented in \citet{vac03}.  
All observations were reduced using Spextool \citep{cus04}.  These 
data were taken during a period of extremely poor weather conditions on Mauna Kea, 
thus we have not attempted to perform a rigorous flux calibration.  While these observing conditions did introduce small artifacts in wavelength regimes 
outside of those presented here (see \citealt{wis05}), we are fully confident that the line profile morphologies and line-to-continuum ratios of the data presented in this paper are not influenced by the presence of any residual telluric features. 

The optical spectroscopic data presented here were obtained at the 1m telescope of the 
Ritter Observatory, using its fiber-fed 
echelle spectrograph.  Nine non-overlapping orders, each of width 70 \AA, were obtained over the wavelength range 
5300-6600\AA\ at R $\sim$26000.  Standard IRAF\footnote{IRAF
is distributed by the National Optical Astronomy Observatories, which are
operated by the Association of Universities for Research in Astronomy, Inc.,
under contract with the National Science Foundation.} techniques were used to reduce these data, including 
procedures to apply bias, flat field, and wavelength calibration corrections to all data.  Further details regarding 
the reduction of Ritter data can be found in \citet{mor97}.  A summary of the optical and IR data to be discussed 
in this paper are presented in Table \ref{photsummary}.  Note that we have not corrected any of our optical or IR 
spectral line data for underlying photospheric absorption components.

\section{Results and Discussion}

\begin{figure}
\centering
\includegraphics[width=8cm]{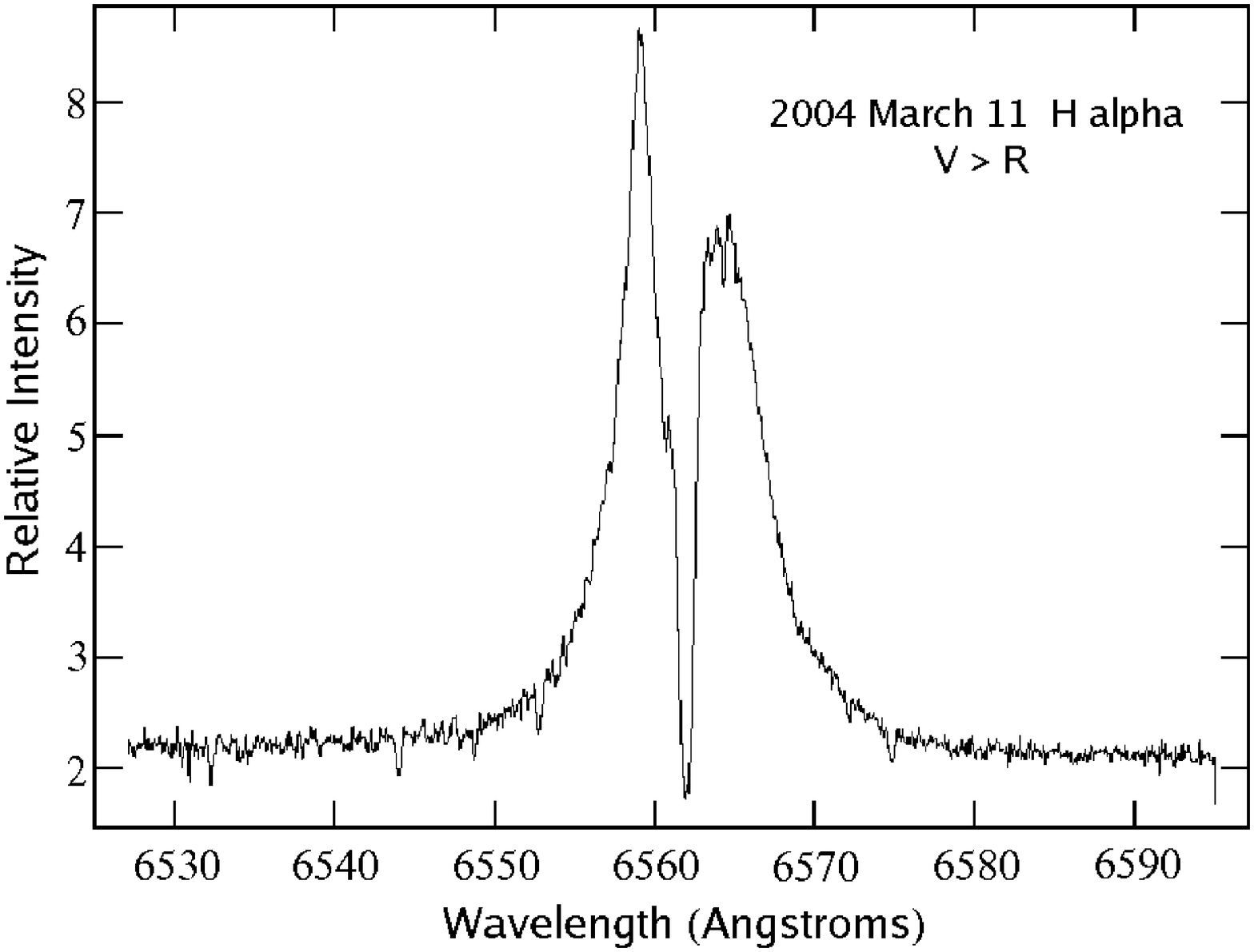}
\includegraphics[width=8cm]{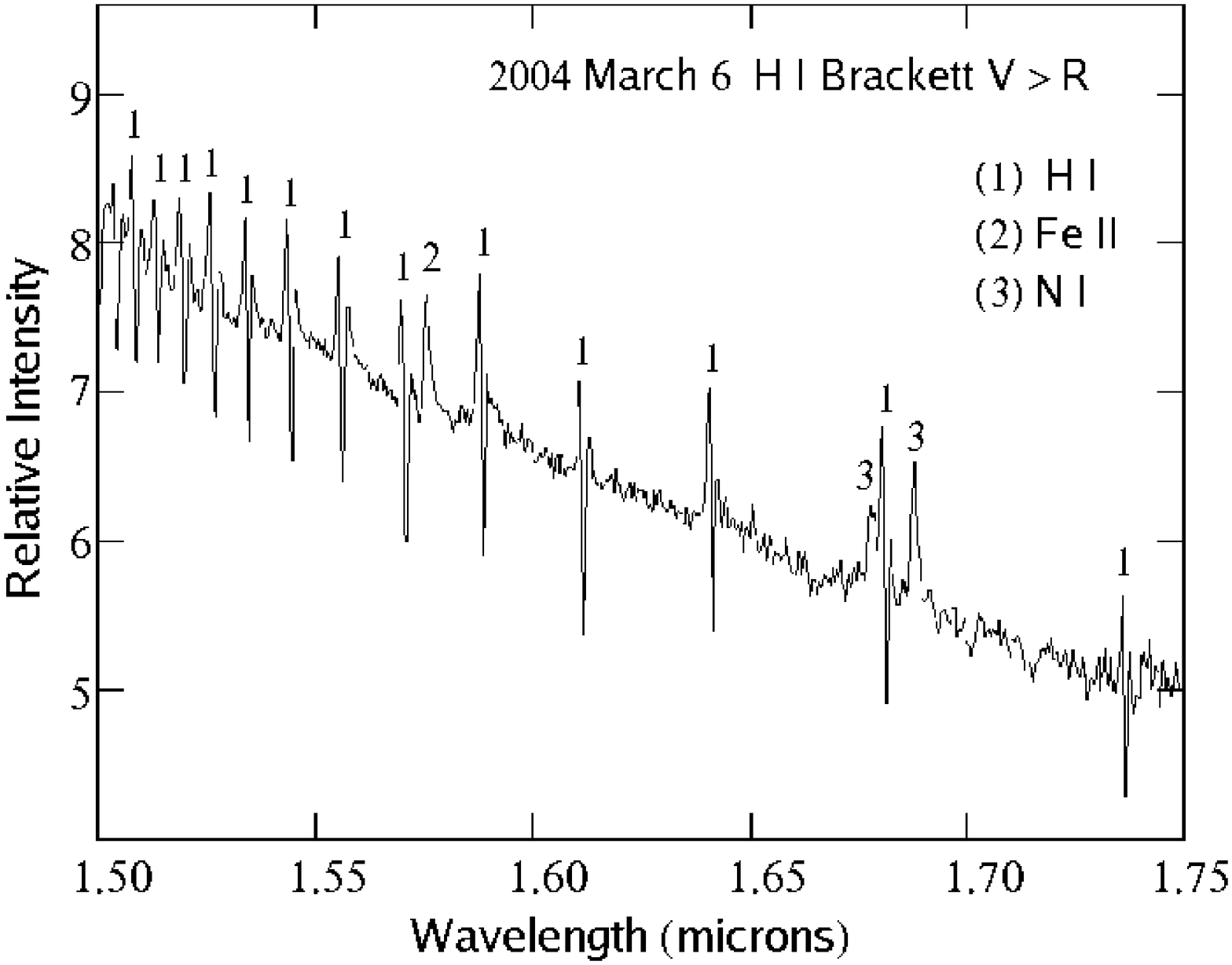}
\caption{\textit{Left:} The H$\alpha$ spectrum of 48 Librae observed on 2004 March 11 
exhibits an asymmetrical profile characterized by V $>$ R.  \textit{Right:} A near contemporaneous 
infrared spectrum obtained on 2004 March 6 also exhibits hydrogen Brackett lines having 
asymmetrical profiles with V $>$ R.  The agreement between the orientation of these line 
profile morphologies is the standard scenario expected in observations of classical Be stars. \label{48lib}}
\end{figure}

As demonstrated by our near-contemporaneous optical and IR observations of 48 Librae 
(see Figure \ref{48lib}), double-peaked 
asymmetrical profiles in Be spectra often exhibit the same V/R orientation for all observed transitions.    
\citet{sle82} reported phase lags between optical Fe II and hydrogen 
lines in several of his observations of Be stars; we found evidence of a similar phenomenon 
in our IR observation of the 
suggested classical Be star NGC 2439:WBBe1 (see \citealt{wis05,wi06b}).  

\begin{figure}
\centering
\includegraphics[width=8cm]{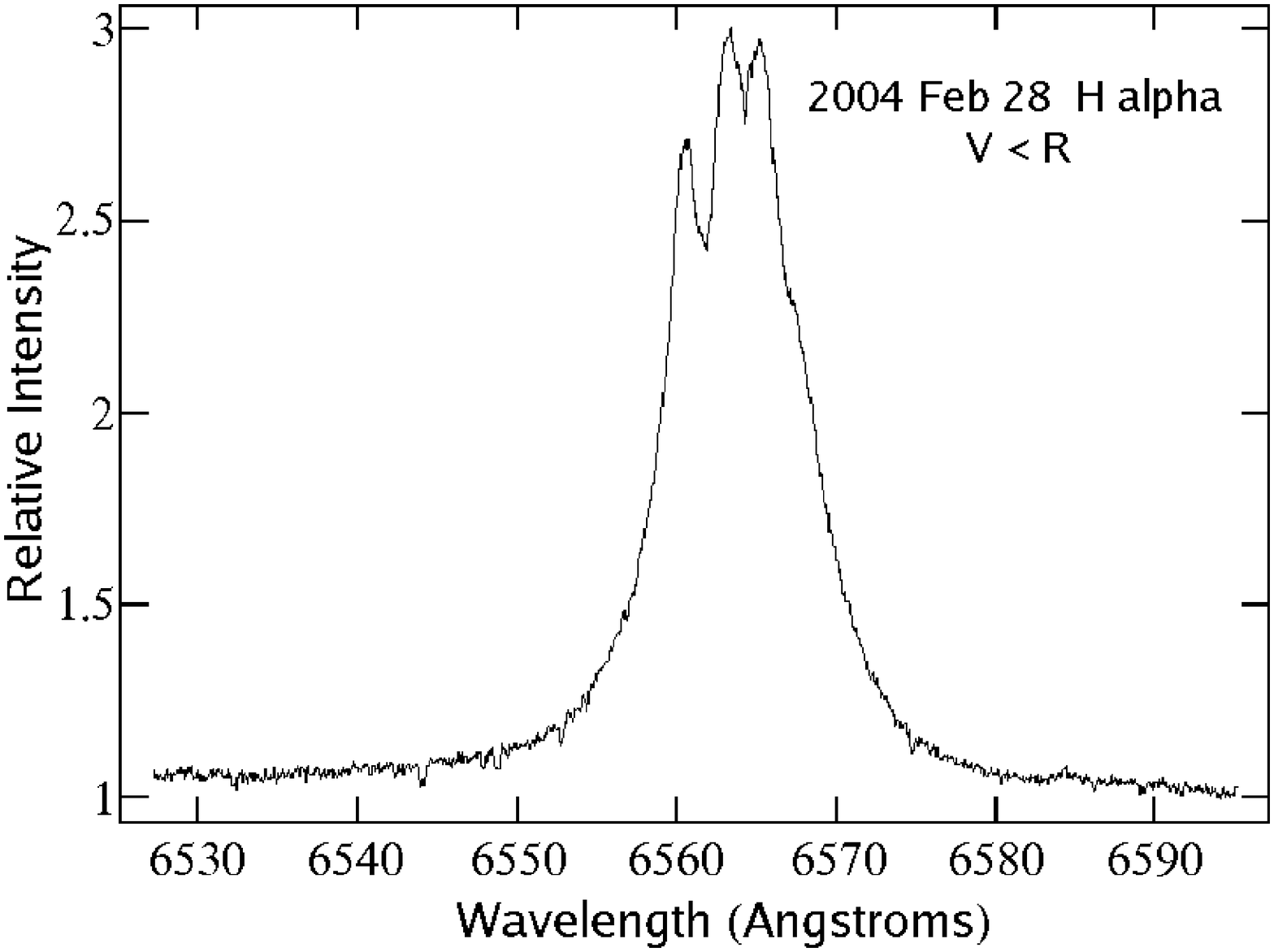}
\includegraphics[width=7.5cm]{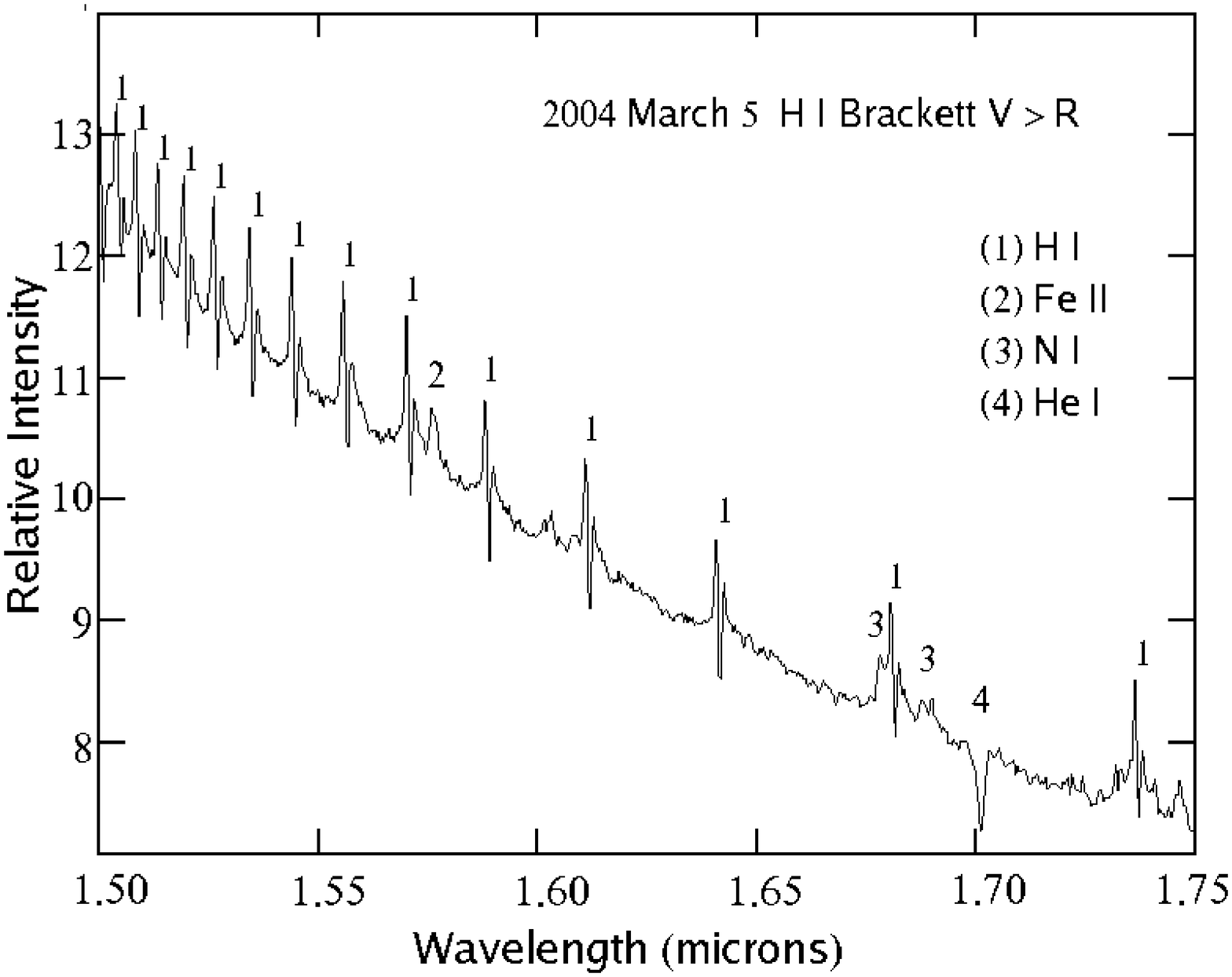}
\caption{\textit{Left:} The H$\alpha$ spectrum of $\zeta$ Tau obtained on 2004 February 28 (left panel) clearly exhibits 
an asymmetrical triple-peaked profile characterized by V $<$ R.  Although not shown here, 
an optical spectrum obtained on 
2004 April 12 exhibits the same morphology.  \textit{Right:} This IR spectrum of $\zeta$ Tau 
obtained on 2004 March 5 is characterized by asymmetrical hydrogen Brackett lines having V $>$ R.  Although not picture here, additional He I, O I, 
Fe II, and H I (Paschen and Pfund series) emission lines in this spectrum exhibit this same V $>$ R morphology. 
\label{ztauoptical}}
\end{figure}

We obtained two H$\alpha$ spectra of the well studied classical Be star $\zeta$ Tau (see Figure \ref{ztauoptical}) one month before and one month after our IR observations, and note that both H$\alpha$ profiles exhibit triple-peaked 
profiles characterized by an overall V $<$ R asymmetry.  The V/R period is often several 
years, so the line profiles are generally stable on timescales of several months \citep{oka97}.  
We are confident that $\zeta$ Tau exhibited a H$\alpha$ 
profile similar to that shown in Figure \ref{ztauoptical} at the time of our IR observation, and was not 
in the midst of a cycle of irregular fluctuations of the type described by \citet{guo95}, based 
on our long baseline of previous observations of the star.  At the resolution of SpeX, $\zeta$ Tau's IR spectrum shows many emission-lines characterized by a range of morphologies, including single-peaked profiles, 
asymmetrical double-peaked profiles, and asymmetrical double-peaked shell profiles.  As shown in 
Figure \ref{ztauoptical}, hydrogen Brackett transitions in these data exhibit shell-like profiles, which
 are asymmetrical with V $>$ R.  Other IR emission-lines which 
have asymmetrical profiles with a similar V/R orientation to the H I Brackett series 
include O I 0.8446, 1.129, 1.317 $\mu$m; Fe II 1.086, 1.113 $\mu$m; He I 1.083 $\mu$m; H I Paschen $\gamma$ and 
$\beta$; and the H I Pfund lines at 2.382, 2.392, 2.403 $\mu$m.  These data exhibit a fundamentally different V/R 
morphology than our near-contemporaneous H$\alpha$ profile.

\begin{figure}
\centering
\includegraphics[width=8cm]{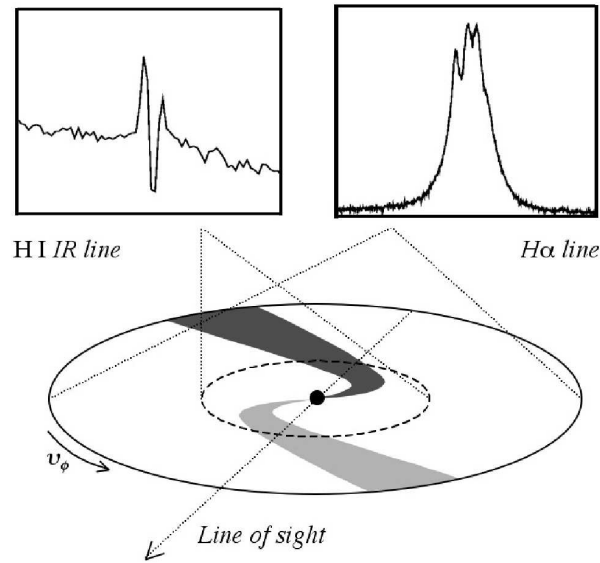}
\caption{A one-armed density perturbation, characterized by regions of density enhancement (lighter regions) 
and regions of density decrements (darker regions), is believed to reside in $\zeta$ Tau's disk based on 
observed V/R line profile variations.  Our IR (left panel) and optical (right panel) 
spectroscopic data probe different 
physical locations within the $\zeta$ Tau disk, hence probe different regions of the density perturbation 
located within the disk. \label{diagram}}
\end{figure}

As $\zeta$ Tau is known to experience cyclical V/R variations, likely due to the presence of a density wave in its disk \citep{oka97}, we suggest that the observed opposite V/R orientation of the IR lines and H$\alpha$ can be explained by the presence of a density wave in the disk which has a significantly different average azimuthal morphology (e.g. a spiral wave), combined with different loci for line formation.
This idea is illustrated in Figure \ref{diagram}, where we show schematically the disk of $\zeta$ Tau with a one-armed spiral density perturbation: note that lighter shading denotes regions of density enhancement while 
darker shading denotes density decrements.  The H I IR lines are much more optically thin than the optical lines, hence they form within the inner disk, while the optical H I lines  originate from a much larger area. As a result of the different azimuthal properties of the disk with respect to the distance from the star, in certain phases of the V/R cycle the net density enhancement of the inner disk may occur in an opposite side (with respect to the line of sight) to the net density enhancement of the entire disk, thus producing line profiles with opposite V/R ratios. 
For instance, in Figure \ref{diagram} the V $>$ R morphology of Br-12 is produced by a net density enhancement in the inner disk region which is moving towards the observer, and the V $<$ R morphology of H$\alpha$ is produced by a net density enhancement in the outer disk region which is moving away from the observer.

\begin{figure}
\centering
\includegraphics[width=8cm]{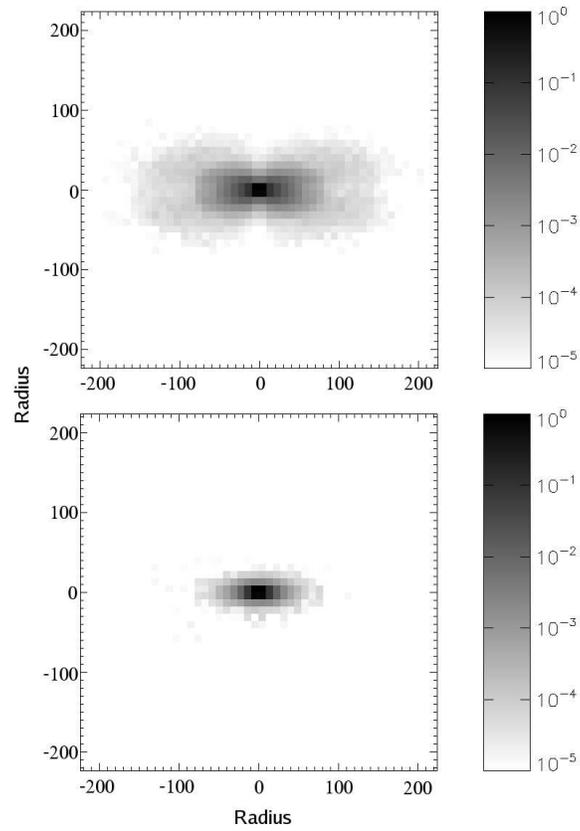}
\caption{\citet{car06} model images of the $\zeta$ Tau disk at H$\alpha$ (upper panel) and H I Br-12 (lower panel).  The scale 
of these images is 40 stellar radii ($\sim$224 solar radii).  These 
images clearly demonstrate that H$\alpha$ is produced over a much larger disk area, whereas H I Br-12 is produced 
from a more inner disk region. \label{allmodel}}
\end{figure}

\citet{car06}, \citet{ca06b}, and \citet{car07} outline a new Monte Carlo code for solving the radiative transfer in the circumstellar disks of Be stars. This fully three-dimensional non-LTE Monte Carlo code 
produces a self-consistent determination of the hydrogen level populations, electron temperature, and gas density of a circumstellar disk, given a prescription of the central star and assuming a static, nearly-Keplerian disk with a given mass loss rate, truncation radius, and a prescription for the gas viscosity.
In \citet{car05}, the authors used this code to study the circumstellar disk of $\zeta$ Tau. Following \citet{woo97}, they adopted 
 a luminosity class of IV and a spectral type of $\rm B2.9 \pm 0.4$, which corresponds to an effective temperature of $19,000 \rm K$ and a stellar radius of $5.6 R_{\sun}$. Assuming an outer disk radius of $100\;R_\star$ and a mass loss rate of $2.5 \times 10^{-11}\;M_{\sun}/\rm yr$ the authors were able to fit the observed SED from visible to mid IR wavelengths, and optical ($\lambda\lambda$ 3200--10500 \AA) spectropolarimetry.
Using this model, we calculated intensity maps of H$\alpha$ and Br-12 for $\zeta$ Tau which, as seen 
in Figure \ref{allmodel}, illustrate that Br-12 is formed in the inner disk ($R \lesssim 15R_\star$) whereas H$\alpha$ is produced over a much more extended region, thereby confirming our previous assertion 
that our optical versus IR H I data probe distinctly different disk regions.  Future iterations of the model will 
incorporate complex density profiles, such as one-armed waves, which will enable us to test our exploratory 
discussion in a more quantitatve manner. 

We note that V/R reversals have been predicted by at least one other theory in the 
refereed literature.  \citet{wat92} predicted instances in which a ratio of R$>$V in a H$\alpha$ line would reverse to a ratio of R$<$V in IR H I lines, 
due to line self-absorption in the inner region of edge-on disks.  However, the presence of this reversal depends critically on the existence of a large expansion component in the disk velocity at the location where the lines 
form, an unlikely scenario given the very small radial velocities now believed to characterize inner disk regions  \citep{riv99,han00,por03}.

\section{Summary and Future Work}

We have presented the first results of near-contemporaneous optical and IR spectroscopic observations of  
classical Be stars.  We 
report the discovery of asymmetrical infrared emission lines 
(He I, O I, Fe II, H I Paschen, Brackett, and Pfund series) in $\zeta$ Tau which have oppositely 
oriented V/R ratios to that  
seen at H$\alpha$.  We suggest these data indicate that the density wave which 
characterizes $\zeta$ Tau's disk has 
a significantly different average azimuthal morphology in the inner disk region as compared to its outer disk region.  This would also indicate that the Fe II lines are formed in the inner disk, in the same spatial location as the IR 
lines are formed.  These features should travel through the disk on the cyclical timescale 
generally reported for density waves in $\zeta$ Tau \citep{oka97}; hence, we would predict continued 
anti-correlation amongst these line profiles over the course of several observing seasons.  We have begun 
a long-term optical and IR spectroscopic monitoring program to trace this evolution.  As demonstrated 
by the initial results presented in this Letter, such an observational approach has the potential to provide 
powerful constraints to theoretical efforts to map the detailed density distribution of Be circumstellar disks, 
hence better understand the mechanism(s) which create these disks.

\acknowledgements 
We thank our referee, Myron Smith, for comments which helped to improve this
manuscript.  We also thank the 
staff of Ritter Observatory for their assistance in providing data presented in this paper.  
This research was supported by a NASA LTSA 
grant to KSB (NAG5-8054), a Research Corporation Cottrell Scholar award (KSB), 
a NASA GSRP Fellowship to JPW (NGT5-50469), a Sigma Xi grant to JPW, and the NSF REU program at the University of Toledo (AFK).  JEB and ACC acknowledge support from NSF 
grant AST-0307686; ACC is also supported by FAPESP grants 01/12589-1 and 04/07707-3.  Observations 
at the Ritter Observatory are supported by 
the NSF under the PREST program, grant number AST-0440784.


\clearpage

\clearpage

\clearpage

\clearpage

\clearpage

\begin{table}
\centering
\begin{center}
\caption{Summary of Observations \label{photsummary}} 
\begin{tabular}{lccc}
\tableline\tableline
Object & Observatory & Date & Exposure Times \\
\tableline
$\zeta$ Tau & I & 2004 March 5 & 60 s. \\
$\zeta$ Tau & R & 2004 February 28 & 1200 s. \\
$\zeta$ Tau & R & 2004 April 12 & 1200 s. \\
48 Lib & I & 2004 March 6 & 150 s. \\ 
48 Lib & R & 2004 March 11 & 3600 s. \\
\tableline
\tablecomments{Some properties of our observational data are summarized.  The observatory abbreviation 
``I'' corresponds to IRTF data and ``R'' corresponds to Ritter data.  For our infrared observations, the listed 
exposure times are the total effective integration times of the science targets, and do not represent the 
individual exposure times at each of the telescope nod positions described in the text.}
\end{tabular}
\end{center}
\end{table}

\end{document}